\documentclass[aps,prx,twocolumn,superscriptaddress,longbibliography]{revtex4-1}
\usepackage{amsfonts}
\usepackage{amsmath}
\usepackage{mathtools}
\usepackage{longtable}
\usepackage{amssymb}
\usepackage{graphicx}
\usepackage{bm}
\usepackage{color}
\usepackage{mathrsfs}
\usepackage[colorlinks,bookmarks=true,citecolor=blue,linkcolor=red,urlcolor=blue]{hyperref}
\usepackage{appendix}
\usepackage{float}
\usepackage{booktabs}
\usepackage[export]{adjustbox}

\newcommand{\RNum}[1]{\uppercase\expandafter{\romannumeral #1\relax}}

\begin{document}
\title{Superdiffusive transport on lattices with nodal impurities}

\author{Yu-Peng Wang}
\thanks{These authors contributed to this work equally.}
\affiliation{Beijing National Laboratory for Condensed Matter Physics and Institute of Physics, Chinese Academy of Sciences, Beijing 100190, China}
\affiliation{University of Chinese Academy of Sciences, Beijing 100049, China}
\author{Jie Ren}
\thanks{These authors contributed to this work equally.}
\affiliation{Beijing National Laboratory for Condensed Matter Physics and Institute of Physics, Chinese Academy of Sciences, Beijing 100190, China}
\affiliation{University of Chinese Academy of Sciences, Beijing 100049, China}
\author{Chen Fang}
\email{cfang@iphy.ac.cn}
\affiliation{Beijing National Laboratory for Condensed Matter Physics and Institute of Physics, Chinese Academy of Sciences, Beijing 100190, China}
\affiliation{Songshan Lake Materials Laboratory, Dongguan, Guangdong 523808, China}
\affiliation{Kavli Institute for Theoretical Sciences, Chinese Academy of Sciences, Beijing 100190, China}

\begin{abstract}
	We show that 1D lattice models exhibit superdiffusive transport in the presence of random ``nodal impurities'' in the absence of interaction.
	Here a nodal impurity is defined as a localized state, the wave function of which has zeros (nodes) in the momentum representation.
	The dynamics exponent $z$, a defining quantity for transport behaviors, is computed to establish this result.
	To be specific, in a disordered system having only nodal impurities, the dynamical exponent $z=4n/(4n-1)$, where $n$ is the order of the node.
	If the system has time reversal, the nodes appear in pairs and the dynamical exponent can be enhanced to $z=8n/(8n-1)$.
	As $1<z<2$, both cases indicate superdiffusive transport.
\end{abstract}

\maketitle

\section{Introduction}
Transport properties unveil fundamental characteristics in quantum systems \cite{transport1, transport2, transport3, heat1, heat2, heat3, info1, info2, info3, info4, info5, info6}. Depending on how energy, charge, or other conserved quantities propagate, transport manifests in three typical categories: localized, diffusive, and ballistic. General chaotic systems typically exhibit diffusive behavior \cite{transport1}, while many integrable systems showcase ballistic transport owing to the presence of non-decaying quasiparticles \cite{transport1}. In quadratic systems with random impurities, the inability of local conserved charges to propagate results in system localization \cite{anderson1958,Thouless1972,hirota1971}.

The transport dynamics are characterized by the dynamical exponent $z$, representing how the width $\sigma$ of the wave packet spreads in time, defined as $\sigma \sim t^{1/z}$ for the late-time limit. The values of $z$ correspond to distinct transport classes: $z=1,2,\infty$ signify ballistic, diffusive, and localized transport, respectively. Superdiffusive transport, indicated by $1<z<2$, is considered anomalous, often involving special underlying mechanisms \cite{superdiffusion1}.

In non-interacting systems, certain types of aperiodic or correlated disorder cause superdiffusive transport, as in the random dimer model \cite{randdimmer} and the Fibonacci model \cite{Fibonacci1,Fibonacci2,Fibonacci3}. Additionally, harmonic chains with random mass \cite{Harmonic1971,Dhar2001} or random magnetic field \cite{Cane2021} also exhibit superdiffusive transport.

Recently, superdiffusive transport is proposed in the one-dimensional spin-1/2 Heisenberg model with the dynamical exponent of $z=3/2$, for the first time in interacting models\cite{znidaric11,prosen17,prosen19,GHD0,GHD00,GHD1,GHD3,GHD4,GHD5}.
It is further predicted that all integrable models with non-Abelian symmetries exhibit $z=3/2$ superdiffusive transport\cite{GHD2,Norman22}. 
Additionally, superdiffusive energy transport has been observed in kinetically constrained models, although a clear analytic understanding of this phenomenon is still lacking \cite{papic23}.
Moreover, long-range interactions, breaking locality, also have the potential to induce superdiffusive transport\cite{longrange0,longrange1,longrange2,longrange3,longrange4,Zhou2020}.

In a recent proposal\cite{wang23}, superdiffusion is induced by a special ``dephasing'' in open quantum systems.
Typically, dephasing appears systems in which the particle density $\hat c^\dagger_i \hat{c}_i$ couples to an environment without memory.
A free fermion subject to such dephasing exhibits diffusive behavior \cite{esposito2005,znidaric_2010,Cao19}.
However, superdiffusive transport appears if the particle $\hat{c}_i$ is replaced by the quasiparticle $\hat{d}_i$, where
\begin{equation}\label{eq:d}
\hat{d}_i\equiv\sum_x d_x\hat{c}_{i+x}
\end{equation}
satisfies that (i) $d_x$ be local, i.e., $d_x=0$ for $|x|>R$, and (ii) its Fourier transform $d_k=\sum_x{d}_xe^{ikx}$ have at least one zero at some $k_0$.
Intuitively, the Bloch waves near $k_0$ have small dephasing probability and hence long lifetimes in propagation, causing the superdiffusion.

In this Letter, we revisit one of the most well-studied problems in transport, the localization problem on 1D lattices with random impurities in the absence of interaction.
We adopt a modified version of the 1D Anderson model
\begin{equation}\label{eq:model}
    \hat H = \hat H_0+\hat V= \sum_{k}E_k\hat{c}^\dag_k\hat{c}_k + \sum_i \epsilon_i\hat{d}_i^\dagger\hat{d}_i,
\end{equation}
where $E_k$ is the band dispersion, and $\epsilon_i$ are the random energies of impurity states satisfying $\overline{\epsilon_i\epsilon_j}=W\delta_{ij}$, $W$ being the disorder strength.
The only modification, inspired by Ref.~\cite{wang23}, is that we have replaced the onsite impurity state $\hat{c}_i$ with $\hat{d}_i$ as defined in Eq.~\ref{eq:d}.
We call $\hat{d}_i$ a nodal impurity if the wave function in $k$-space, $d_{k_0}=0$, for some $k_0$.

In the original Anderson model, the system shows absence of transport at all energies for arbitrarily small impurity strength: this is called the Anderson localization\cite{anderson1958,Thouless1972,hirota1971}.
We show that the above modification $\hat{c}_i\rightarrow\hat{d}_i$ leads to superdiffusive transport, if $\hat{d}_i$'s are nodal impurities.
To be specific, we prove that $z=4n/(4n-1)$, where $n$ is the order of zero at the node $k_0$.
In realistic systems, time-reversal symmetry is common, and it dictates that the nodes at $\pm{k_0}$ appear in pairs.
When only $E_{-k_0}$ is degenerate with $E_{k_0}$, we show that
time-reversal symmetry enhances the superdiffusive transport, resulting in $z=8n/(8n-1)$.
These findings are corroborated by extensive numerical results in large systems. 

\section{Model and analysis}
The eigenfunction of the free Hamiltonian $\hat H_0=\sum_k E_k \hat{c}_k^\dagger \hat{c}_k$ are plane waves, represented by $|\psi_k\rangle = \sum_x e^{ikx}\hat{c}_x^\dagger |0\rangle$, which exhibits ballistic transport. 
We note that nodal impurities $\hat V=\sum_i \epsilon_i \hat{d}_i^\dagger \hat{d}_i$ satisfies
\begin{equation}
	\langle\psi_k|\hat d^\dagger_i \hat d_i|\psi_k\rangle = \left|\hat d_i|\psi_k\rangle\right|^2 = |d_k|^2,
\end{equation} 
which implies $\hat V|\psi_{k_0}\rangle = 0$ at the node $k_0$. The plane wave eigenstate $|\psi_{k_0}\rangle$ remains unscattered by nodal impurities, leading to the divergence of the localization length $\xi_k$ at $k_0$.

In 1D disordered systems, the transmission probability $T_k$ typically decreases exponentially with system size $L$ \cite{anderson1958}. We can approximately represent it as $T_k=\exp(-L/\xi_k)$, where $\xi_k$ is the localization length.

Now, we want to calculate the transmission probability $T_k$.
First, we examine the single impurity case, assuming the impurity spans $l$ sites, and $\hat{d} = \sum_{a=1}^l d_a \hat{c}_{a}$. The Hamiltonian can be expressed as:
\begin{equation}
	\hat H_{1}= \hat H_0 + \hat V_1 =\sum_k E_k \hat c^\dagger_k c_k + W\hat{d}^\dagger \hat d,
\end{equation}
where $W$ is impurity strength.

We rewrite this Hamiltonian in momentum space as:
\begin{equation}
	\hat H_1 =\hat H_0 + \hat V_1 = \sum_k E_k \hat c^\dagger_k c_k + W\sum_{k',k}d_{k'}^* d_k \hat c^\dagger_{k'} \hat c_k.
\end{equation}

For simplicity, we assume that every momentum $k$ has only one equal energy partner $k'$ with an opposite velocity. Considering the incoming wave is a plane wave $|k\rangle$, the scattering probability to the equal-energy outgoing wave to $|k'\rangle$ is (see Appendix.~\ref{app:reflection})
\begin{equation}
R_k^{(1)} = c_k|d_k|^2|d_{k'}|^2
\end{equation}
where $c_k = W^2 v_{k'}^{-2} \left(1 - \sum_{k''} \frac{W |d_{k''}|^2}{\epsilon_{k} - \epsilon_{k''}}\right)^{-2}$ and $v_k = \frac{\mathrm{d}E_k}{\mathrm{d}k}$. In one dimension, $R_1$ is also called the reflection probability. The transmission probability of a single impurity is given by $T_k^{(1)} = 1 - \frac{v_{k'}}{v_k}R_k^{(1)}$.

Without additional symmetry, $d_{k'}$ and $d_k$ would not simultaneously be zero except for specific fine-tuned cases. Thus, near the nodal point $k_0$, $R_k^{(1)}$ behaves as:
\begin{equation}\label{eq:R_nonsysm}
R_{k_0+q}^{(1)} \sim |d_{k_0+q}|^2 \sim q^{2n},
\end{equation}
where $n$ represents the order of the zero at the node $k_0$ of $d_k$.

Next, we consider the multiple-impurity case. Assuming that $\epsilon_i$ for every site $i$ is randomly and independently chosen to be $W$ with probability $p$, or $0$ with a probability of $1-p$. In the thermodynamic limit ($L\to+\infty$), there are, on average, $pL$ impurities in this disordered chain. When the condition $k/p \gg 2\pi$ is satisfied, we have $\lim_{L\to \infty}\log T_k^{(L)}=pL \log T_k^{(1)}$ (see Appendix~\ref{app:logT}), where $T_k^{(L)}$ is the transmission probability of the disordered chain with system size $L$. Therefore, near the node $k_0$, the localization length diverges as:
\begin{equation}\label{eq:loc_div}
\xi_{k_0+q}=-\frac{1}{p\log(1-\frac{v_{k'}}{v_k}R_{k_0+q}^{(1)})} \sim \frac{1}{q^{2n}}
\end{equation}

\begin{figure*}
	\centering
	\includegraphics[width=\linewidth]{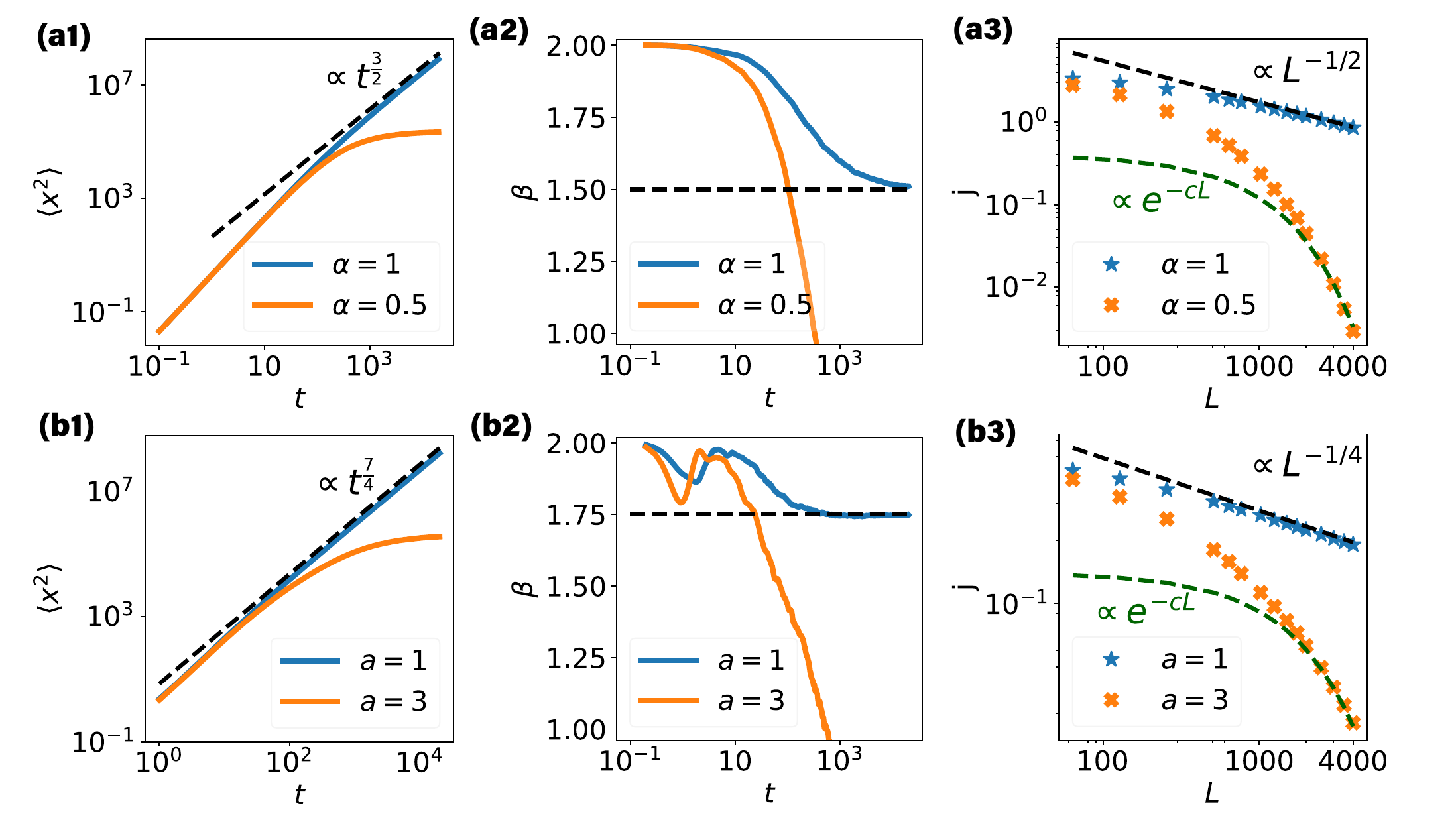}
	\caption{
	\textbf{Numerical results of nodal impurity models:}
	(a1)(a2)(a3) for 2-site inpurity $\hat{d}_i = (\hat{c}_i + i\alpha\hat{c}_{i+1})/\sqrt{1+|\alpha|^2}$, and (b1)(b2)(b3) for 3-site time-reversal impurity $\hat{d}_i = (\hat{c}_{i-1} + a\hat{c}_i + \hat c_{i+1})/\sqrt{2+a^2}$.
	(a1) \textbf{Evolution of mean-square displacement in 2-site impurity model:}
	Numerical simulations depict the evolution of mean-square displacement in 2-site impurity model $\hat{d}_i = (\hat{c}_{i} + \alpha i\hat{c}_{i+1})/\sqrt{1+|\alpha|^2}$, showcasing nodal impurity with $\alpha=1$ (blue line) and non-nodal impurity with $\alpha=0.5$ (orange line). The initial state is $\psi(x)=\delta_{0,x}$. The impurity chain with nodal impurities exhibits superdiffusive transport, characterized by an asymptotic behavior of $\langle x^2 \rangle \sim t^{3/2}$ (black dashed line), while the non-nodal impurity model displays localized behavior.
	(b1) \textbf{Evolution of mean-square displacement in 3-site disorder model:}
	The same for 3-site impurity $\hat{d}_i = (\hat{c}_{i-1} + a\hat{c}_i + \hat c_{i+1})/\sqrt{2+a^2}$. Blue line represents the nodal case $a=1$, and the orange line represents the non-nodal case $a=3$.
	(a2) \textbf{Exponent evolution in 2-site impurity model:}
	Depicts the evolution of the exponent $\beta = \frac{d \log \langle x^2 \rangle(t)}{d\log t}$ for the model with 2-site impurity $\hat{d}_i = (\hat{c}_{i} + i\hat{c}_{i+1})/\sqrt{2}$. The behavior of $\beta(t)$ converges to $3/2$ at long times.
	(b2) \textbf{Exponent evolution in 3-site impurity model:}
	Depicts the evolution of the same exponent for the model with 3-site disorder $\hat{d}_i = (\hat{c}_{i-1} + \hat{c}_i + \hat c_{i+1})/\sqrt{3}$. The behavior of $\beta(t)$ converges to $7/4$ at long times.
	(a3) \textbf{NESS current in 2-site impurity model:}
	NESS current of a boundary-driven quasi-particle disorder chain with 2-site impurity $\hat{d}_i = \hat{c}_{i} + \alpha i\hat{c}_{i+1}$, featuring nodal impurities with $\alpha=1$ (blue line) and non-nodal impurities with $\alpha=0.5$ (orange line). The current of the nodal disorder chain decreases as $j\sim L^{-1/2}$, whereas the non-nodal disorder chain exhibits exponential decay.
	(b3) \textbf{NESS current in 3-site impurity model:}
	The same for 3-site impurity $\hat{d}_i = (\hat{c}_{i-1} + a\hat{c}_i +	\hat c_{i+1})/\sqrt{2+a^2}$. The blue line represents the nodal case $a=1$, whose current decreases as $j\sim L^{-1/4}$. The orange line represents the non-nodal case $a=3$, whose current decreases exponentially.
	In all the figures, the disorder strength is $W=1$, and $\epsilon_i$ for every site $i$ is randomly and independently chosen to be $W$ with a probability of $p=1/20$, or $0$ with a probability of $1-p$. All the data is averaged over 200 random disorder configurations.
	}
	\label{fig:fig1}
\end{figure*}

We now illustrate how the divergence of the localization length leads to superdiffusive transport and determines the dynamical exponent. Initially, we explore the current of the nonequilibrium steady state (NESS) under boundary driving\cite{boundarydrivenRMP,znidaric16,znidaric16,Prosen_2009,znidaric11}.

We couple the first and last sites to baths described phenomenologically by the following four Lindblad operators:
\begin{equation}\label{eq:bd_driven}
\begin{aligned}
	\mathcal{L}^{(\text{bath})}(\hat \rho)=\sum_{m=1}^4 2 \hat L_m \hat \rho \hat L_m^{\dagger}&-\hat \rho L_m^{ \dagger} \hat L_m-\hat L_m^{ \dagger} \hat L_m \hat \rho \\
	\hat L_1=\sqrt{\Gamma(1+\mu)} \hat c^\dagger_1,\quad & \hat L_2=\sqrt{\Gamma(1-\mu)} \hat c_1 \\
	\hat L_3=\sqrt{\Gamma(1-\mu)} \hat c^\dagger_L,\quad & \hat L_4=\sqrt{\Gamma(1+\mu)} \hat c_L
\end{aligned}
\end{equation}
where $\hat \rho$ represents the density matrix. The density matrix's evolution follows the Lindblad master equation:
\begin{equation}\label{eq:Lindblad}
	\frac{\mathrm{d} \hat \rho}{\mathrm{d} t}=\mathrm{i}[\hat \rho, \hat H]+\mathcal{L}^{(\text {bath})}(\hat \rho).
\end{equation}
According to Ref.~\cite{Thierry2020}, the NESS current can be obtained as:
\begin{equation}
\begin{aligned}
	j &= \mu\Gamma \int \mathrm{d}k\ |v_k| T_k \\&\overset{L\to\infty}{\longrightarrow} \mu\Gamma \int \mathrm{d}k\ |v_k| e^{-L/\xi_k},
\end{aligned}
\end{equation}
where $v_k = \frac{\mathrm{d} E_k}{\mathrm{d} k}$.

In the thermodynamic limit $L\to \infty$, this integral is dominated by the momenta near node $k_0$. Given the divergent behavior of the localization length near $k_0$ (Eq.~\ref{eq:loc_div}) and $v_{k_0}\ne 0$, we have:
\begin{equation}
	j \sim \int \mathrm{d}q\ e^{-pL q^{2n}} \sim L^{-1/2n}
\end{equation}

Under this boundary-driven condition (Eq.~\ref{eq:bd_driven}), difference of particle number between two boundaries $\delta \tilde n$ is fixed. Considering Fick's law $j = D\frac{d\tilde n(x)}{dx} = D \frac{\delta \tilde{n}}{L}$, $j \sim L^{-1/2n}$ implies the scaling of the diffusion constant $D\sim L^{1-1/2n}$. Given that the states contributing to transport traverse the system's length with a finite constant velocity ($v_{k_0}\ne 0$), the terms $t$ and $L$ can be interchanged, resulting in $D\sim t^{1-1/2n}$. Consequently, this leads to the superdiffusive behavior $\langle x^2\rangle \sim Dt \sim t^{\frac{4n-1}{2n}}$, and $z=4n/(4n-1)$.

It's noteworthy that while previous studies assert that in generic quantum systems, $j\sim L^{-\gamma}$ implies $z=1+\gamma$\cite{znidaric16,boundarydrivenRMP}, our model presents a unique scenario. In generic cases, most states contribute to the transport. However, in our case, only states with momenta near the node contribute to transport. All states contributing to transport possess similar velocities, while other states remain localized. Therefore, the relationship $j\sim L^{-\gamma}$ in our model implies $z=2/(2-\gamma)$. In both cases, $\gamma=1$ (Ohm's law) indicates diffusion, $\gamma = 0$ (current is independent of system size) indicates ballistic transport, and $0<\gamma<1$ indicates superdiffusive transport.

Time-reversal symmetry plays a crucial role in shaping the dynamics of the system. Under this symmetry, the energy dispersion of $\hat H_0$ satisfies $E_k=E_{-k}$, and the Fourier transform of $\hat d_i$'s wave function satisfies $d_k=d_{-k}^*$. In general, there always exists a momentum region $\bm{K}$ such that only $-k$ has the same energy as momentum $k$. In other word, when $k\in \bm{K}$, $E_{k'}\ne E_k$ unless $k'=\pm k$.

Within this region $\bm{K}$, the reflection probability of a single impurity can be expressed as:
\begin{equation}
	R_k^{(1)} = c_k|d_k|^2|d_{-k}|^2.
\end{equation}
Moreover, if $d_{k_0}=0$, it implies $d_{-k_0}=0$. If $k_0 \in \bm{K}$, this doubles the order of zeros of the reflection probability, leading to an enhanced divergent behavior of the localization length near node $k_0$:
\begin{equation}
	\xi_{k_0+q} = 1/q^{4n}.
\end{equation}

Consequently, the dynamical exponent is given by $z=\frac{8n}{8n-1}$, and $\langle x^2 \rangle\sim t^{2-1/4n}$.

\section{numerics}

Next, we present numerical simulations for specific models to verify the above analysis.
The simplest nodal impurity model is a nearest-neighbor hopping with 2-site impurities:
\begin{equation}\label{eq:Model}
	\hat H = \sum_i (\hat c_i^\dagger \hat c_{i+1}+ \hat c_{i+1}^\dagger\hat c_i) + \sum_i \epsilon_i \hat d_i^\dagger \hat d_i
\end{equation}
where $\hat d_i = (\hat c_{i}+ \alpha e^{i\theta} \hat c_{i+1})/\sqrt{1+|\alpha|^2}$, $\alpha$ represents a positive real number, and $\theta$ ranges from $0$ to $2\pi$. When $\alpha = 1$, the momentum distribution $|d_k|^2 = (1+\alpha e^{i(\theta+k)})/\sqrt{1+|\alpha|^2}$ exhibits a nodal point at $k_0 = \pi -\theta$ with order $n=1$. Conversely, when $\alpha\ne1$, such as $\alpha=0.5$, the momentum distribution exhibits no nodal points.

A single-particle state can be represented by $\Psi(t) = \sum_x \phi_j(t) \hat{c}_j^\dagger |0\rangle $. Since this model corresponds to a quadratic free fermion, we investigate its dynamic exponent $z$ through the evolution of the point initial state $\phi_x(t=0) = \delta_{0,j}$. Notably, the mean square displacement asymptotically grows as $\langle x^2 \rangle \sim t^{2/z}$, where $\langle x^2 \rangle(t) \equiv \sum_j j^2 |\phi_j(t)|^2$. 

In our numerical simulation, we set $\theta=\pi/2$. For $\alpha = 1$, $d_k$ has a node at $k = \pi/2$ with order $n=1$. The mean square displacement asymptotically grows as $\langle x^2 \rangle \sim t^{3/2}$(Fig.~\ref{fig:fig1}(a1)(a2)) and the current under fixed boundary driven scales with system size as $j\sim L^{-1/2}$(Fig.~\ref{fig:fig1}(a3)), implying a dynamic exponent of $z= \frac {4}{3}$, characterizing the system as superdiffusion. For $\alpha=0.5$, $d_k$ has no nodal points, resulting in the system behaving similarly to the Anderson model, indicating localization(Fig.~\ref{fig:fig1}(a1)(a3)). The details of the boundary-driven model setup are provided in Appendix.~\ref{app:boundary driven}.

In Eq.~\ref{eq:Model}, TRS is broken by the nodal impurity as $d_k\ne d_{-k}^*$. The simplest TRS invariant nodal impurity involves three sites: $\hat d_i = (\hat c_{i-1}+a\hat c_i+\hat c_{i+1})/\sqrt{(2+a^2)}$, where $a$ is a real number. Its Fourier transform is given by $d_k = (2\cos k+a)/\sqrt{2+a^2}$. In this case, $\bm{K}=[-\pi,\pi]$ is the whole Brillouin zone.

For $|a|<2$, $d_k$ has nodes with order $n=1$. Numerical simulations reveal that the nonequilibrium steady-state (NESS) current of this disordered chain under boundary-driven conditions scales with the system size as $j\sim L^{-1/4}$ (Fig.~\ref{fig:fig1}(b3)), and the mean square displacement grows as $\langle x^2\rangle \sim t^{7/4}$ at late times (Fig.~\ref{fig:fig1}(b1)(b2)).

In contrast, under condition $|a|>2$, $d_k$ exhibits no nodes. Numerical simulations indicate that the NESS current of this disordered chain under boundary-driven conditions decreases exponentially (Fig.~\ref{fig:fig1}(b3)), and the evolution of mean square displacement suggests system localization (Fig.~\ref{fig:fig1}(b1)).

While the analytical results assume that $\epsilon_i$ for every site $i$ is randomly and independently chosen to be $W$ with probability $p$, or $0$ with a probability of $1-p$, our numerical simulations demonstrate that randomly sampling $\epsilon_i$ from the range $(-W/2, W/2)$ with uniform probability yields similar results (see Appendix.~\ref{app:numeric}).

\section{conclusion and discussion}

In this paper, we present a new mechanism that induces superdiffusive transport in quadratic disordered systems. This mechanism requires two key elements:
\begin{enumerate}
\item \textbf{Free fermions Hamiltonian} which possesses ballistic eigenmodes.
\item \textbf{Nodal impurities} that localize most eigenmodes, except for a measure-zero set of eigenmodes with nodal momentum $k_0$.
\end{enumerate}
These two elements result in a power-law divergence of the localization length $\xi(k)$ at the node $k_0$. Modes with momenta near the nodes contribute to the superdiffusive transport, with the dynamical exponent $z$ determined by the order $n$ of the node, given by $z = (4n - 1) / 4n$. Additionally, this superdiffusive behavior is enhanced under time-reversal symmetry, resulting in a dynamical exponent of $z = (8n - 1) / 8n$.

It is crucial to note that the concept of ``nodal points" is not limited to disordered systems; similar phenomena are observed in dephasing systems \cite{wang23,znidaric24}. The underlying philosophy of this ``nodal points" picture is both simple and general. The first element, free fermions, can be replaced by integrable models, which also possesses ballistic eigenmodes. The second element, nodal impurities, can be replaced by dephasing or interaction. This approach allows for the construction of many superdiffusive models in different systems.

For example, we can replace the free fermion Hamiltonian $\hat{H}_0$ in our model with an integrable model that possesses ballistic modes. As long as we can find some local operator $O_i$ that scatters these ballistic modes to diffusive or localized modes but leaves a specific mode with momentum $k_0$ unscattered, we can construct a superdiffusive model using these operators as impurities, dephasing, or interactions. This approach may lead to the discovery of chaotic models exhibiting superdiffusive transport, a topic we leave for future study.

\begin{acknowledgements}

\end{acknowledgements}
Y.-P. W. thanks Marko \v Znidari\v c for his valuable comments.

\begin{appendix}

	\begin{widetext}
		\section{Reflection probability}\label{app:reflection}
		Consider the Hamiltonian
		\begin{equation}\label{eq:1}
		H=H_0+V=\sum_k \epsilon_k|k\rangle\langle{k}|+\sum_{k,k'}Wd_kd^\ast_{k'}|k\rangle\langle{k'}|.
		\end{equation}
		Now we assume an eigenstate $|\psi_k\rangle$ given in the iterative expression
		\begin{equation}\label{eq:2}
		|\psi_k\rangle=|k\rangle+\sum_{k'}|k'\rangle\frac{\langle{k'}|V|\psi_k\rangle}{\epsilon_k-\epsilon_{k'}}.
		\end{equation}
		We define the T-matrix
		\begin{equation}\label{eq:3}
		T_{k'k}=\langle{k'}|V|\psi_k\rangle=\langle{k'}|V|k\rangle+\sum_{k''}\frac{\langle{k'}|V|k''\rangle{T}_{k''k}}{\epsilon_k-\epsilon_{k''}}.
		\end{equation}
		As
		\begin{equation}\label{eq:4}
		V_{k'k}=Wd_k'd^\ast_k,
		\end{equation}
		we postulate
		\begin{equation}\label{eq:5}
		T_{k'k}=Wd_k'd^\ast_kt_k.
		\end{equation}
		Substituting Eqs.(\ref{eq:4},\ref{eq:5}) into Eq.(\ref{eq:3}), we have
		\begin{equation}\label{eq:6}
		t_k=\frac{1}{1-\sum_{k''}\frac{W|d_{k''}|^2}{\epsilon_{k}-\epsilon_{k''}}}.
		\end{equation}
		Substituting Eq.(\ref{eq:6}) into Eq.(\ref{eq:2}), we have
		\begin{equation}
		|\psi_k\rangle=|k\rangle+\frac{Wd^\ast_k}{1-\sum_{k''}\frac{W|d_{k''}|^2}{\epsilon_k-\epsilon_{k''}}}\sum_{k'}\frac{d_k}{\epsilon_k-\epsilon_{k'}}|k'\rangle.
		\end{equation}
		Consider the wavefunction in the position representation
		\begin{equation}
		\psi_k(x)=e^{ikx}-Wt_kd^\ast_k\sum_{k'}\frac{d_{k'}e^{ik'x}}{\epsilon_{k'}-\epsilon_{k}}.
		\end{equation}
		Now we use a trick by adding a small imaginary part to the denominator in the summation
		\begin{equation}
		\sum_{k'}\frac{d_{k'}e^{ik'x}}{\epsilon_{k'}-\epsilon_{k}}\rightarrow\sum_{k'}\frac{d_{k'}e^{ik'x}}{\epsilon_{k'}-\epsilon_{k}+i\delta}
		\end{equation}
		Now we change the variable
		\begin{equation}
		z=e^{ik'},
		\end{equation}
		such that
		\begin{eqnarray}
		\sum_k'&\rightarrow&\frac{1}{2\pi}\int{dk'}\rightarrow\frac{1}{2i\pi}\oint\frac{dz}{z},\\
		\nonumber
		\epsilon_{k'}&\rightarrow&\epsilon(z),\\
		\nonumber
		e^{ik'x}&\rightarrow&{z}^x,\\
		\nonumber
		d_{k'}&\rightarrow&{d}(z),\\
		\nonumber
		\frac{d}{dk}&\rightarrow&{iz}\frac{d}{dz}\\
		\label{eq:12}
		\sum_{k'}\frac{d_{k'}e^{ik'x}}{\epsilon_{k'}-\epsilon_{k}+i\delta}&\rightarrow&\frac{1}{2\pi{i}}\oint{dz}\frac{d(z)z^{x-1}}{\epsilon(z)-\epsilon_k-i\delta}.
		\end{eqnarray}
		$d(z),\epsilon(z)$ are all meromorphic functions, that is, have poles and zeros at a finite number of points.
		Therefore, we can use the residue theorem to evaluate the integral in Eq.(\ref{eq:12}).
		
		Consider three loops: loop-a is the unit circle which is the loop of integration in Eq.(\ref{eq:12}); loop--b is an infinitesimal loop that contains no poles of the integrand other than the origin; loop-c is a large circle which keeps all the poles of the integrand outside except $\infty$.
		Loop-a is defined as counterclockwise, and loop-b,c are defined as clockwise.
		From Riemann's theorem of integration, we have
		\begin{eqnarray}\label{eq:13}
		\oint_{a}+\oint_b&=&(2\pi{i})\sum_{p_i\;\mathrm{between}\;a\;and\;b}R_i,\\
		\label{eq:14}
		\oint_{a}+\oint_c&=&-(2\pi{i})\sum_{p_i\;\mathrm{between}\;a\;and\;c}R_i.
		\end{eqnarray}
		For $x>0$ and $x$ large enough, due to the factor $z^{x-1}$, we have
		\begin{equation}
		\oint_b=0,
		\end{equation}
		so using Eq.(\ref{eq:13})
		\begin{equation}
		\frac{1}{2\pi{i}}\oint{dz}\frac{d(z)z^{x-1}}{\epsilon(z)-\epsilon_k-i\delta}=\sum_i\frac{d(z_i)z_i^{x-1}}{d\epsilon(z_i)/dz}.
		\end{equation}
		Here $z_i$'s are the zeros of $\epsilon(z)-\epsilon_k-i\delta=0$ inside the unit circle.
		Among all the zeros, there is a special zero
		\begin{equation}
		\zeta=e^{ik}+\frac{i\delta}{d\epsilon'(e^{ik})/dz}+O(\delta^2)=e^{ik}(1-v_k^{-1}\delta)+O(\delta^2)
		\end{equation}
		that has the largest $|z_i|$, so as $x\rightarrow+\infty$, this is the only pole that survives (assuming $v_k>0$)
		\begin{equation}
		\psi_k(x\rightarrow+\infty)=e^{ikx}+iWt_ke^{ikx}|d_k|^2v_k^{-1}.
		\end{equation}
		Similarly for $x<0$ and $|x|$ sufficiently large, we have
		\begin{equation}
		\oint_c=0,
		\end{equation}
		so using Eq.(\ref{eq:14})
		\begin{equation}
		\frac{1}{2\pi{i}}\oint{dz}\frac{d(z)z^{x-1}}{\epsilon(z)-\epsilon_k-i\delta}=-\sum_i\frac{d(z_i)z_i^{x-1}}{d\epsilon(z_i)/dz}.
		\end{equation}
		where $z_i$'s are zeros of $\epsilon(z)-\epsilon_k-i\delta=0$ outside the unit circle.
		Among all the zeros, there is a special zero
		\begin{equation}
		\zeta_r=e^{ik_r}+\frac{i\delta}{d\epsilon'(e^{ik_r})/dz}+O(\delta^2)=e^{ik}(1-v_{k_r}^{-1}\delta)+O(\delta^2),
		\end{equation}
		where $\epsilon_{k_r}=\epsilon_k$.
		$k_r$ represents the momentum of the reflected wave, so if $v_k>0$, then $v_{k_r}<0$, so $\zeta_r$ is indeed outside the unit circle.
		After some simple derivation, we have
		\begin{equation}
		\psi_k(x\rightarrow-\infty)=e^{ikx}+iWt_ke^{-ik_rx}d^\ast_kd_{k_r}|v_{k_r}|^{-1}.
		\end{equation}
		Therefore the reflection probability
		\begin{equation}
		R=|Wt_kd_k^\ast{d}_{k_r}|^2/v_{k_r}^2.
		\end{equation}
		
		\section{Transmission probability with multiple impurities}\label{app:logT}
		In this appendix, following the demonstration of \cite{muller2010}, we show that the averaged logarithm transmission probability of multiple impurities $\langle\log T\rangle$ is just the sum of logarithm transmission probability of single impurity when impurity density is low. 
		
		At first, we consider the scattering process of single impurity $V = \hat d^\dagger\hat d$, where $\hat d = \sum_{a=1}^l d_a \hat c_a$. We can decompose the wave function at the left ($L, x<0$) and right ($R,x>l$) into incoming and outgoing waves:
		\begin{align}
			\phi_L(x) = \phi^{in}_L e^{ikx}+\phi^{out}_L e^{ik'x}\\
			\phi_R(x) = \phi^{out}_R e^{ikx}+\phi^{in}_R e^{ik'x}
		\end{align}
		The outgoing amplitudes are linked to the incident amplitudes by the reflection and transmission coefficients $r$ and $t$ from the left, and $r^\prime$, $t^\prime$ from the right:
		\begin{align}
			\phi^{out}_L = t\phi^{in}_L + r^\prime \phi^{in}_R\\
			\phi^{out}_R = r\phi^{in}_L + t^\prime \phi^{in}_R
		\end{align}
		
		The scattering matrix can be defined as 
		\begin{align}\label{eq:Smatrix}
			\begin{pmatrix}
				\phi_L^{out}\\ \phi_R^{out}
			\end{pmatrix} = 
			S \begin{pmatrix}
				\phi_L^{in}\\ \phi_R^{in}
			\end{pmatrix}
			\quad \text{where}\ S = \begin{pmatrix}
				r &t^\prime \\
				t &r^\prime 
			\end{pmatrix}.
		\end{align}
		The reflection and transmission probability from the left are respectively $R=|r|^2$ and $T=|t|^2$, and similar from the right $R^\prime=|r^\prime|^2$ and $T^\prime=|t^\prime|^2$.
		The probability flux conservation ensure that $S$ is unitary, $S^\dagger S = Id$, which leads to $T=T^\prime$, $R=R^\prime$ and $T+R=T^\prime+R^\prime=1$.
		
		We can also decompose the wave function into left-moving and right-moving components:
		\begin{align}
			\phi_L(x) = \phi^{+}_L e^{ikx}+\phi^{-}_L e^{ik'x}\\
			\phi_R(x) = \phi^{+}_R e^{ikx}+\phi^{-}_R e^{ik'x}.
		\end{align}
		The transfer matrix $M$ maps the the amplitudes from the left side of this impurity to the right:
		\begin{align}
			\begin{pmatrix}
				\phi_R^{+}\\ \phi_R^{-}
			\end{pmatrix} = 
			M \begin{pmatrix}
				\phi_L^{+}\\ \phi_L^{-}
			\end{pmatrix}.
		\end{align}
		From equation~(\ref{eq:Smatrix}), we can get the transfer matrix
		\begin{align}
			M = \begin{pmatrix}
				t-rr^\prime/t^\prime &r^\prime /t^\prime \\
				-r/t & 1/t
			\end{pmatrix}.
		\end{align}
		
		In multiple impurities case, the total transfer matrix is just the product of transfer matrixes of all impurities. In two impurity case, $M_{12} = M_{1}M_2$. The total transmission amplitude is 
		\begin{equation}
			t_{12} = \frac{t_1 t_2}{1-r_1^\prime r_2}.
		\end{equation}
		The logarithm transmission probability is 
		\begin{align}
			\log T_{12} = \log T_1 +\log T_2 + \log |1-\sqrt{R_1R_2}e^{i\theta}|,
		\end{align}
		where $\theta$ is the total phase accumulated during one complete internal reflection.
		
		Considering that $\epsilon_i$ in inpurity model (Eq.~\ref{eq:model}) for every site $i$ is randomly and independently chosen to be $W$ with probability $p$, or $0$ with a probability of $1-p$.
		\begin{align}\label{eq:model}
			\hat H &= \sum_k E_k \hat c_k^\dagger \hat c_k + \sum_i \epsilon_i \hat d_i^\dagger \hat d_i
		\end{align}
		There are average $pL$ impurities in this disorder chain. Under the condition $k/p \gg 2\pi$, the log-averaged transmission probability satisfies
		\begin{equation}
			\log T_k^{(L)}= pL\log T_k^{(1)} + \sum_{i=1}^{pL} \log |1-R_k^{(1)} e^{i\theta_i}|,
		\end{equation}
		where $\theta_i$ is the total phase accumulated during one complete internal reflection between the $i$th and $(i+1)$th impurities.
		
		When distance $\delta x$ between two impurities is randomly and $ k \delta x\gg 2\pi$, $\theta_i$ is also randomly distributed in $[0,2\pi]$. We have
		\begin{equation}
			\lim_{pL\to \infty}\sum_{i=1}^{pL} \log |1-R_k^{(1)} e^{i\theta_i}| = pL \int_0^{2\pi} d\theta \log |1-R_k^{(1)}e^{i\theta}|=0,
		\end{equation}
		and
		\begin{equation}
			\lim_{pL\to \infty}\log T_k^{(L)}= pL\log T_k^{(1)}
		\end{equation}
		
		Therefore, the localization length can be represented as $\xi(k) = -1/(pL\log T_1(k))$.

		\section{Setup of boundary-driven model} \label{app:boundary driven}
		In this appendix, we illustrate the boundary driven setup utilized to derive the scaling relation between NESS current and system size, a method commonly employed to investigate transport properties in various studies\cite{znidaric11,znidaric16,znidaric20,znidaric17}.
		
		We couple the first and the last site to baths described phenomenologically by the following 4 Lindblad operators,
		\begin{align}
			\mathcal{L}^{(\text{bath})}(\rho)=\sum_{k=1}^4 2 L_k \rho L_k^{\dagger}&-\rho L_k^{ \dagger} L_k-L_k^{ \dagger} L_k \rho \\
			L_1=\sqrt{\Gamma(1+\mu)} \hat c^\dagger_1,\quad & L_2=\sqrt{\Gamma(1-\mu)} \hat c_1 \\
			L_3=\sqrt{\Gamma(1-\mu)} \hat c^\dagger_L,\quad & L_4=\sqrt{\Gamma(1+\mu)} \hat c_L
		\end{align}
		where $\rho$ is density matrix. The density matrix's evolution is governed by the Lindblad master equation:
		\begin{equation}\label{eq:Lindblad}
			\frac{\mathrm{d} \rho}{\mathrm{d} t}=\mathrm{i}[\rho, \hat H]+\mathcal{L}^{(\text {bath})}(\rho).
		\end{equation}
		Here, $\hat H= \sum_i (\hat c_i^\dagger \hat c_{i+1}+ \hat c_{i+1}^\dagger\hat c_i) + \sum_i \epsilon_i \hat d_i^\dagger \hat d_i = \sum_{ij}H_{ij}\hat c^\dagger_i \hat c_j $ represents the Hamiltonian of the impurity model. In free fermion case, current $j$ is proportional to $\mu$. Without loss of generality, we set $\Gamma=\mu=1$. As the complete Liouvillean is quadratic, the equation of motion is closed, and the NESS is specified by the two-point green function $C$, where $C_{ij}=\langle \hat c^\dagger_i \hat c_j\rangle = \operatorname{Tr}(\hat c^\dagger_i \hat c_j \rho)$.
		
		Since the density matrix satisfies the Lindblad master equation (\ref{eq:Lindblad}), in the Heisenberg picture, the evolution of an operator $\hat O$ satisfies:
		\begin{align}\label{eq:operator_evo}
			\frac{\mathrm{d}\hat O}{\mathrm{d} t} &= i[\hat H,\hat O] + \sum_k 2L_k^\dagger \hat O L_k - \hat O L^\dagger_k L_k - L^\dagger_k L_k \hat O\\
			&=i[\hat H,\hat O] + \sum_k L_k^\dagger [\hat O,L_k] + [L^\dagger_k, \hat O] L_k.
		\end{align}
		
		By substituting $\hat O =\hat c^\dagger_i \hat c_j$ into Eq. (\ref{eq:operator_evo}), we obtain the evolution of the two-point green function:
		\begin{equation}
			\frac{\mathrm{d} C}{\mathrm{d} t} = XC+CX^\dagger+P,
		\end{equation}
		where $X=iH^T-R$, $R_{11}=R_{LL} = 1$, $P_{11}=2$, and all other elements of $R$ and $P$ is zero.
		
		The current of $\hat H_0 = \sum_i (\hat c_i^\dagger \hat c_{i+1}+ \hat c_{i+1}^\dagger\hat c_i)$ is 
		\begin{align}
			j=\langle \hat c_i^\dagger\hat c_{i+1}- \hat c_{i+1}^\dagger\hat c_{i}\rangle = 2\operatorname{Im}C_{i,i+1}.
		\end{align}
		Considering the case of an l-site disorder $\hat d_i = \sum_{a=0}^{l-1}d_{a} \hat c_{i+a} $, we let $\epsilon_1=\epsilon_2=0$, ensuring no disorder hopping between sites $1$ and $2$. Consequently, the NESS current of the disorder chain is $j = 2\operatorname{Im}C_{1,2}$.

		\section{Numerical results for sampling $\epsilon_i$ from uniform distribution}\label{app:numeric}
		In this section, we present numerical simulations (Fig.~\ref{fig:supfig1}) for the case that disorder strength $\epsilon_i$ is randomly and independently sampled from $[-W/2,W/2]$. The physical quantity is the same as Fig.~\ref{fig:fig1} in main text.
		
		\begin{figure*}
			\centering
			\includegraphics[width=\linewidth]{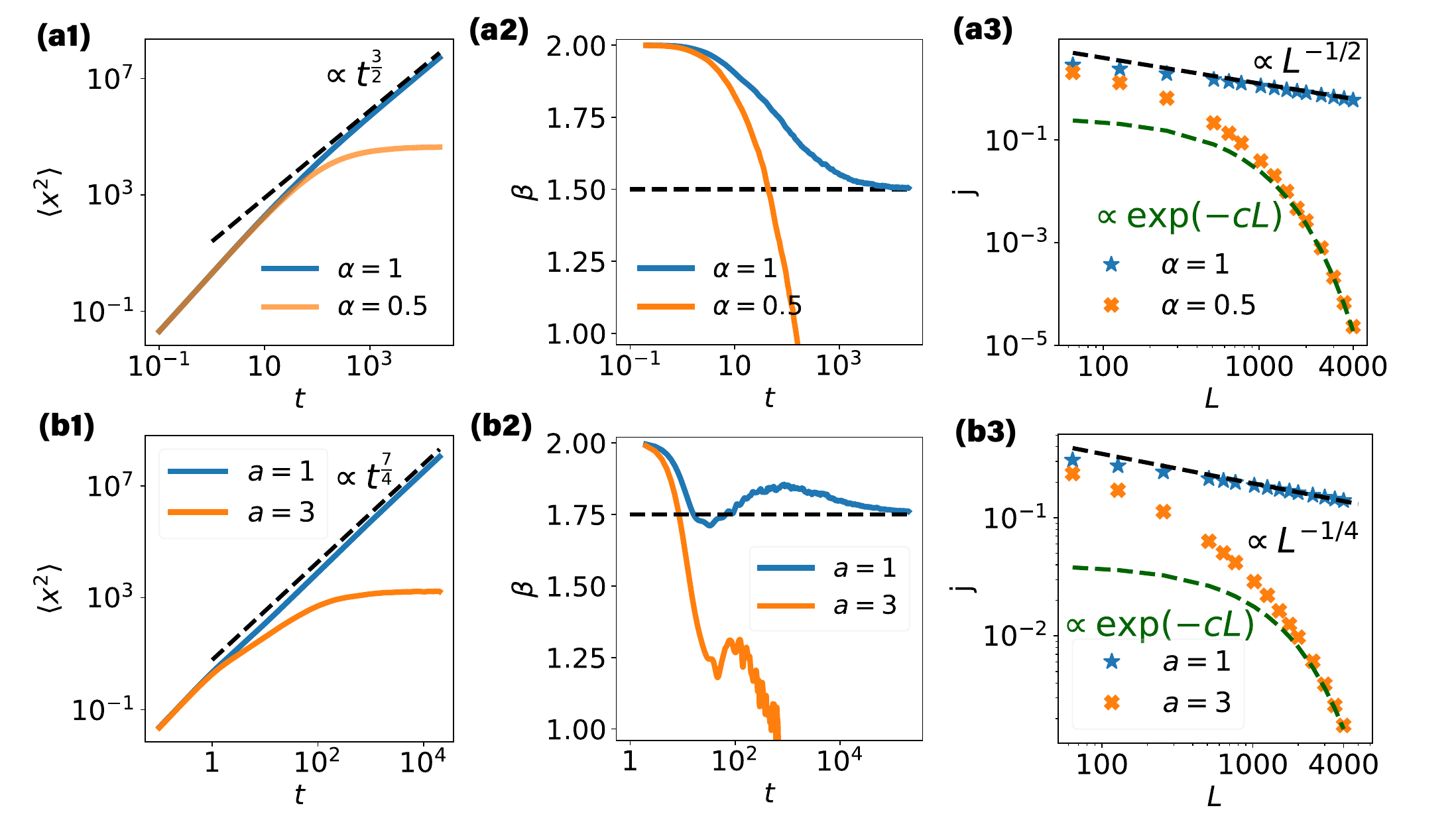}
			\caption{
				\textbf{Numerical results of nodal impurity models:}
				(a1)(a2)(a3) for 2-site inpurity $\hat{d}_i = (\hat{c}_i + i\alpha\hat{c}_{i+1})/\sqrt{1+|\alpha|^2}$, and (b1)(b2)(b3) for 3-site time-reversal impurity $\hat{d}_i = (\hat{c}_{i-1} + a\hat{c}_i + \hat c_{i+1})/\sqrt{2+a^2}$.
				(a1) \textbf{Evolution of mean-square displacement in 2-site impurity model:}
				Numerical simulations depict the evolution of mean-square displacement in 2-site impurity model $\hat{d}_i = (\hat{c}_{i} + \alpha i\hat{c}_{i+1})/\sqrt{1+|\alpha|^2}$, showcasing nodal impurity with $\alpha=1$ (blue line) and non-nodal impurity with $\alpha=0.5$ (orange line). The initial state is $\psi(x)=\delta_{0,x}$. The impurity chain with nodal impurities exhibits superdiffusive transport, characterized by an asymptotic behavior of $\langle x^2 \rangle \sim t^{3/2}$ (black dashed line), while the non-nodal impurity model displays localized behavior.
				(b1) \textbf{Evolution of mean-square displacement in 3-site disorder model:}
				The same for 3-site impurity $\hat{d}_i = (\hat{c}_{i-1} + a\hat{c}_i + \hat c_{i+1})/\sqrt{2+a^2}$. Blue line represents the nodal case $a=1$, and the orange line represents the non-nodal case $a=3$.
				(a2) \textbf{Exponent evolution in 2-site impurity model:}
				Depicts the evolution of the exponent $\beta(t) = \frac{d \log \langle x^2 \rangle(t)}{d\log t}$ for the model with 2-site impurity $\hat{d}_i = (\hat{c}_{i} + i\hat{c}_{i+1})/\sqrt{2}$. The behavior of $\beta(t)$ converges to $3/2$ at long times.
				(b2) \textbf{Exponent evolution in 3-site impurity model:}
				Depicts the evolution of the same exponent for the model with 3-site disorder $\hat{d}_i = (\hat{c}_{i-1} + \hat{c}_i + \hat c_{i+1})/\sqrt{3}$. The behavior of $\beta(t)$ converges to $7/4$ at long times.
				(a3) \textbf{NESS current in 2-site impurity model:}
				NESS current of a boundary-driven quasi-particle disorder chain with 2-site impurity $\hat{d}_i = \hat{c}_{i} + \alpha i\hat{c}_{i+1}$, featuring nodal impurities with $\alpha=1$ (blue line) and non-nodal impurities with $\alpha=0.5$ (orange line). The current of the nodal disorder chain decreases as $j\sim L^{-1/2}$, whereas the non-nodal disorder chain exhibits exponential decay.
				(b3) \textbf{NESS current in 3-site impurity model:}
				The same for 3-site impurity $\hat{d}_i = (\hat{c}_{i-1} + a\hat{c}_i +	\hat c_{i+1})/\sqrt{2+a^2}$. The blue line represents the nodal case $a=1$, whose current decreases as $j\sim L^{-1/4}$. The orange line represents the non-nodal case $a=3$, whose current decreases exponentially.
			In all the figures, the disorder strength is $W=1$, and $\epsilon_i$ for every site $i$ is randomly and independently sampling from $[-W/2,W/2]$. All the data is averaged over 200 random disorder configurations.
			}
			\label{fig:supfig1}
		\end{figure*}

		\end{widetext}

\end{appendix}

\bibliography{ref}

\end{document}